\documentclass[twocolumn,iop]{emulateapj}
\usepackage{apjfonts}

\usepackage{amsmath,graphicx,longtable,hyperref}
\usepackage{natbib,threeparttable,rotating}

\newcommand{\etal}{et al.}
\newcommand{\hbeta}{H{$\beta$}}

\def\FeII{Fe\,{\sc ii}}

\newcommand{\OII}{[O{\sevenrm\,II}]}

\def \OIII {[O\,{\sc iii}]}

\newcommand{\bracket}[1]{\left\langle#1\right\rangle}
   \font\sevenrm=cmr7 scaled 1000

\def\kms{{\rm km\,s^{-1}}}

\slugcomment{ApJ in press}

\begin{document}

\title{The Sloan Digital Sky Survey Reverberation Mapping Project: No Evidence for Evolution in the $M_{\bullet}-\sigma_*$ Relation to $z\sim 1$}


\author{Yue Shen$^{1,2,3}$, Jenny E.~Greene$^{4}$, Luis C.~Ho$^{1,5}$, W.~N. Brandt$^{6,7}$, Kelly D.~Denney$^{8}$, Keith Horne$^9$, Linhua Jiang$^1$, Christopher S.~Kochanek$^{8,10}$, Ian D.~McGreer$^{11}$, Andrea Merloni$^{12}$, Bradley M.~Peterson$^{8,10}$, Patrick Petitjean$^{13}$, Donald P.~Schneider$^{6,7}$, Andreas Schulze$^{14}$, Michael A.~Strauss$^{4}$, Charling Tao$^{15,16}$, Jonathan R.~Trump$^{6,7,3}$, Kaike Pan$^{17}$, Dmitry Bizyaev$^{17}$ }

\altaffiltext{1}{Kavli Institute for Astronomy and Astrophysics, Peking University, Beijing 100871, China}
\altaffiltext{2}{Carnegie Observatories, 813 Santa Barbara Street, Pasadena,
CA 91101, USA}
\altaffiltext{3}{Hubble Fellow}
\altaffiltext{4}{Department of Astrophysical Sciences, Princeton University, Princeton, NJ 08544, USA}
\altaffiltext{5}{Department of Astronomy, School of Physics, Peking University, Beijing 100871, China}
\altaffiltext{6}{Department of Astronomy \& Astrophysics, The Pennsylvania
State University, University Park, PA, 16802, USA}
\altaffiltext{7}{Institute for Gravitation and the Cosmos, The Pennsylvania
State University, University Park, PA 16802, USA}
\altaffiltext{8}{Department of Astronomy, The Ohio State University, 140 West 18th Avenue,
Columbus, OH 43210, USA}
\altaffiltext{9}{SUPA Physics/Astronomy, Univ. of St. Andrews, St. Andrews KY16 9SS, Scotland, UK}
\altaffiltext{10}{Center for Cosmology and AstroParticle Physics, The Ohio State University, 191 West Woodruff Avenue, Columbus, OH 43210, USA}
\altaffiltext{11}{Steward Observatory, The University of Arizona, 933 North Cherry Avenue, Tucson, AZ 85721-0065, USA}
\altaffiltext{12}{Max-Planck Institute fuer Extraterrestrische Physik, Giessenbachstr. 1, 85748 Garching, Germany}
\altaffiltext{13}{Institut d'Astrophysique de Paris, Universit\'e Paris 6 and CNRS, 98bis Boulevard Arago, 75014 Paris, France}
\altaffiltext{14}{Kavli Institute for the Physics and Mathematics of the Universe (Kavli IPMU, WPI), The University of Tokyo, Kashiwa 277-8583, Japan}
\altaffiltext{15}{Centre de Physique des Particules de Marseille, Aix-Marseille Universit\'e , CNRS/IN2P3, 163, avenue de Luminy - Case 902 - 13288 Marseille Cedex 09, France}
\altaffiltext{16}{Tsinghua Center for Astrophysics, Tsinghua University, Beijing 100084, China}
\altaffiltext{17}{Apache Point Observatory and New Mexico State University, P.O. Box 59, Sunspot, NM 88349, USA}

\shorttitle{SDSS-RM: The $M_{\bullet}-\sigma_*$ Relation at High $z$}


\shortauthors{SHEN ET~AL.}

\begin{abstract}

We present host stellar velocity dispersion measurements for a sample of $88$ broad-line quasars at $0.1<z<1$ (46 at $z>0.6$) from the Sloan Digital Sky Survey Reverberation Mapping (SDSS-RM) project. High signal-to-noise ratio coadded spectra (average ${\rm S/N}\approx 30$ per 69~${\rm km\,s^{-1}}$ pixel) from SDSS-RM allowed decomposition of the host and quasar spectra, and measurements of the host stellar velocity dispersions and black hole (BH) masses using the single-epoch (SE) virial method. The large sample size and dynamic range in luminosity  ($L_{5100}=10^{43.2-44.7}\,{\rm erg\,s^{-1}}$) lead to the first clear detection of a correlation between SE virial BH mass and host stellar velocity dispersion far beyond the local universe. However, the observed correlation is significantly flatter than the local relation, suggesting that there are selection biases in high-$z$ luminosity-threshold quasar samples for such studies. Our uniform sample and analysis enable an investigation of the redshift evolution of the $M_{\bullet}-\sigma_*$ relation relatively free of caveats by comparing different samples/analyses at disjoint redshifts. We do not observe evolution of the $M_{\bullet}-\sigma_*$ relation in our sample up to $z\sim 1$, but there is an indication that the relation flattens towards higher redshifts. Coupled with the increasing threshold luminosity with redshift in our sample, this again suggests certain selection biases are at work, and simple simulations demonstrate that a constant $M_\bullet-\sigma_*$ relation is favored to $z\sim 1$. Our results highlight the scientific potential of deep coadded spectroscopy from quasar monitoring programs, and offer a new path to probe the co-evolution of BHs and galaxies at earlier times. 


\end{abstract}


\section{Introduction}\label{sec:intro}

\clearpage

The observed correlations between black hole (BH) mass and host galaxy properties in the local universe have motivated extensive theoretical work on AGN feedback and the co-evolution of supermassive black holes (SMBHs) and their hosts \citep[e.g.,][and references therein]{Benson_2010,Kormendy_Ho_2013}. The $M_\bullet-\sigma_*$ relation between BH mass and bulge stellar velocity dispersion ($\sigma_*$) arguably represents the tightest among these relations, and has been well measured for both inactive galaxies \citep[e.g.,][]{Gebhardt_etal_2000a,Ferrarese_Merritt_2000,Tremaine_etal_2002,Gultekin_etal_2009,McConnell_Ma_2013} and active galaxies \citep[e.g.,][]{Gebhardt_etal_2000b,Ferrarese_etal_2001,Nelson_etal_2004,Onken_etal_2004,Barth_etal_2005,Greene_Ho_2006b,Dasyra_etal_2007,Watson_etal_2008,Grier_etal_2013,Woo_etal_2013, Bennert_etal_2014} in the local universe. 

Given the importance of these BH scaling relations in our understanding of galaxy formation and evolution, there have been extensive observational efforts in the past decade both to refine the measurements of the local relations \citep[e.g.,][and references therein]{Gebhardt_etal_2000a,Ferrarese_Merritt_2000,Graham_etal_2001,Tremaine_etal_2002,
Marconi_Hunt_2003,Haring_Rix_2004,Aller_Richstone_2007, Bentz_etal_2009c, Gultekin_etal_2009, Graham_etal_2011,McConnell_Ma_2013, Kormendy_Ho_2013}, and to probe their possible redshift evolution \citep[e.g.,][]{Treu_etal_2004,Peng_etal_2006a,Peng_etal_2006b,Woo_etal_2006,Treu_etal_2007, Woo_etal_2008,Salviander_etal_2007,Shen_etal_2008b,Jahnke_etal_2009, McLeod_Bechtold_2009,Decarli_etal_2010,Merloni_etal_2010,Bennert_etal_2010,Cisternas_etal_2011,Canalizo_etal_2012,Hiner_etal_2012,Targett_etal_2012,Salviander_Shields_2013, Matsuoka_etal_2014, Bongiorno_etal_2014}. Given the small angular size of the sphere of influence of the BH, so far all measurements beyond the local universe are based on broad-line quasar samples, where the BH mass can be estimated using the properties of the broad lines, instead of spatially resolved gas or stellar kinematics. Most of these studies reported a positive evolution, advocating a scenario where BH growth precedes that of the bulge. However, statistical biases resulting from uncertainties in BH mass estimates and host property measurements, and from the selection biases in broad-line quasar samples with luminosity thresholds provide cautionary notes on such claims \citep[e.g.,][]{Salviander_etal_2007,Lauer_etal_2007b, Shen_etal_2008a, Shen_Kelly_2010, Schulze_Wisotzki_2011, Schulze_Wisotzki_2014, Shen_2013, Salviander_Shields_2013}. 

On the other hand, measuring the host galaxy stellar velocity dispersion of quasars is notoriously difficult. It requires sufficient spectral quality  to measure the host stellar absorption features diluted by the quasar continuum. Such exercises based on survey-quality spectra are limited to low redshifts \citep[e.g.,][]{Greene_Ho_2006a,Greene_Ho_2006b,Shen_etal_2008b}. At $z>0.3$, \citet[][]{Woo_etal_2006} and \citet{Woo_etal_2008} presented $\sigma_*$ measurements in 14 quasars at $z=0.36$ and 5 quasars at $z=0.57$, respectively, based on high S/N Keck spectroscopy. The latter two studies represent the best effort to date in measuring $\sigma_*$ for higher-$z$ quasar hosts. So far there has been no direct measurement of $\sigma_*$ in $z>0.6$ quasars. 

Here we use spectroscopic data from the the Sloan Digital Sky Survey Reverberation Mapping (SDSS-RM) project \citep{Shen_etal_2014} to study the relationship between BH mass and host stellar velocity dispersion in quasars at $0.1<z<1$, a regime where BH masses can only be estimated (other than from BH-galaxy scaling relations) in broad-line quasars with reverberation mapping \citep[e.g.,][]{Peterson_2013} or its extension, the single-epoch virial BH mass estimators \citep[e.g.,][]{Shen_2013}. SDSS-RM is a major multi-object reverberation mapping project; it spectroscopically monitors a flux-limited ($i<21.7$) sample of 849 quasars in a single 7 deg$^2$ field with the SDSS-III BOSS spectrograph \citep[][]{Smee_etal_2013} on the 2.5-m wide-field SDSS telescope \citep{Gunn_etal_2006}. With a total of 32 spectroscopic epochs totaling $\sim 60$ hrs of exposure from the SDSS-RM baseline program in 2014, the coadded spectra provide very high S/N to allow decomposition of the host and quasar spectra, and measurements of BH and host properties. In this work we focus on $\sigma_*$, and more detailed studies of the host stellar populations of these quasars will be presented elsewhere. 

The paper is organized as follows. In \S\ref{sec:data} we describe the data, the host/quasar spectral decomposition, and the measurements of $\sigma_*$ and BH mass. In \S\ref{sec:result} we present the results. We discuss the caveats of our analysis and the interpretation of the data in \S\ref{sec:disc} and conclude in \S\ref{sec:con}. A flat $\Lambda$CDM cosmology is adopted throughout with $\Omega_0=0.3$ and $h_0=0.7$. We use $M_\bullet$ to denote the true BH mass, and $M_{\rm BH,vir}$ to denote the virial BH mass estimate. By default wavelengths are restframe vacuum wavelengths. For ease of discussion, we use ``high-$z$'' to refer to $z>0.3$ in this paper. 

\section{Data and Spectral Measurements}\label{sec:data}

The spectra were taken as part of the SDSS-III BOSS survey \citep[][]{Eisenstein_etal_2011,Dawson_etal_2013} from Jan to Jul 2014. The SDSS-RM field was observed with
an upgraded version of the pair of multi-object fiber spectrographs used for SDSS I/II
\citep{Smee_etal_2013}, with a fiber diameter of 2\arcsec. The wavelength coverage of BOSS spectroscopy is 3650--10,400 \AA, with a spectral resolution of $R\sim 2000$, and a pixel size of $10^{-4}$ in $\log_{10}\lambda$ ($69\,{\rm km\,s^{-1}}$ native SDSS pixel). The typical S/N per pixel in $g$ band in a 2-hr exposure is $\sim 4.5$ at $g_{\rm psf}=21.2$. The spectra for individual epochs were flux-calibrated using a custom approach based on simultaneous standard star observations to an absolute accuracy of $\sim 5\%$ for point sources \citep[][]{Shen_etal_2014}.

All 32 epochs from the 2014 observations of SDSS-RM were coadded using an inverse-variance-weighted spline fit (as in the IDL-based BOSS pipeline\footnote{Publicly available at http://www.sdss3.org/svn/repo/idlspec2d/} {\em idlspec2d}), with a logarithmic wavelength binning of $10^{-4}$ in $\log_{10}\lambda$. The spectra are stored in vacuum wavelength, and calibrated to the heliocentric reference with an accuracy of better than 5 $\kms$. Redshifts were measured using the SDSS redshift pipeline \citep{Bolton_etal_2012} on the coadded spectra and visually verified.

We limit the current study to the 212 $z<1.09$ quasars in the full SDSS-RM sample for which we can decompose the host component from the spectrum, and measure the virial BH mass using the broad \hbeta\ line, arguably the best line to estimate a virial BH mass \citep[see discussion in][Sec 3.1.7]{Shen_2013}. All data and measurements in this paper are provided in a FITS catalog described in Table \ref{table:sample}.

\begin{figure*}
\centering
    \includegraphics[width=0.6\textwidth]{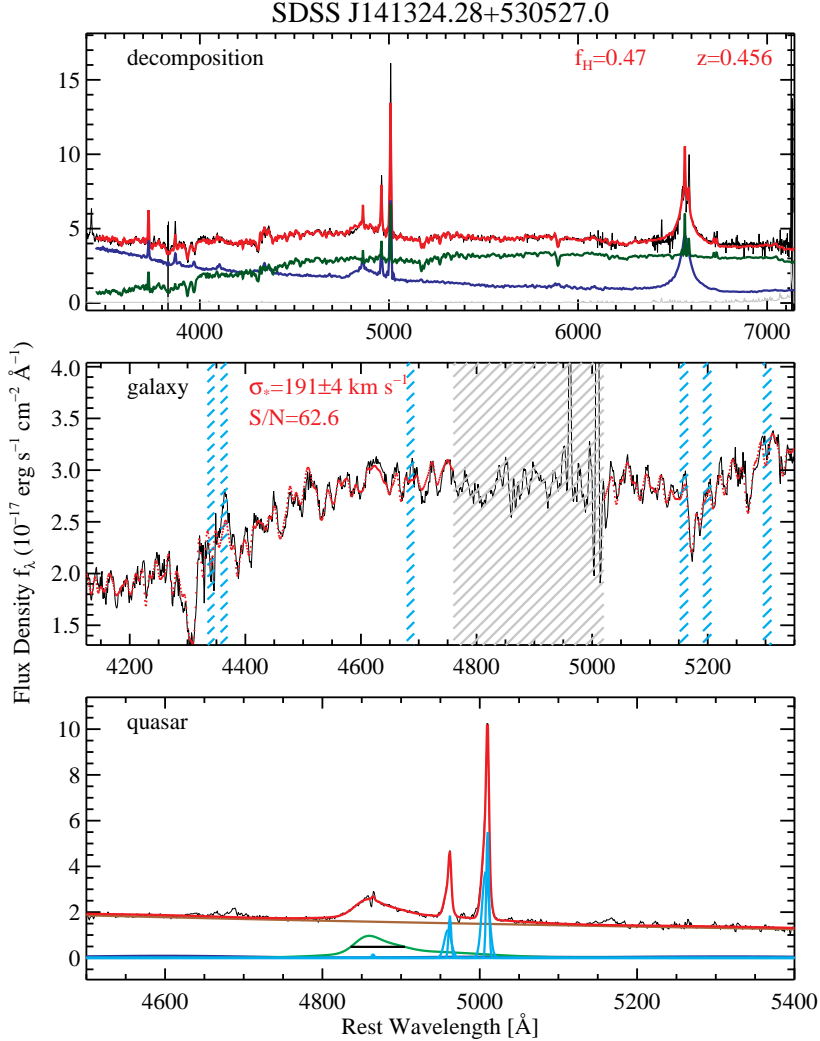}
     \caption{An example of the spectral decomposition (top) and measurements of $\sigma_*$ (middle) and quasar properties (bottom). In the top panel, the estimated host fraction within the wavelength range $4160\,\textrm{\AA}<\lambda<4210\,$\AA\ and the redshift are given in the upper-right corner. The black line is the original spectrum, with spectral errors shown as the gray line. The blue and green lines are the reconstructed quasar and host galaxy spectra, respectively. The red line is the sum of the blue and green lines. There are residuals in the quasar emission line region due to the limited number of quasar eigenspectra used. In the middle panel, the black line is the decomposed galaxy spectrum (original spectrum minus reconstructed quasar spectrum) and the red line is the best-fit from {\em vdispfit}. The gray and cyan shaded bands are regions around quasar broad/narrow emission lines and galaxy narrow emission lines, that were masked for the fits. The best-fit $\sigma_*$ and the median S/N per pixel of the decomposed host spectrum are marked. In the bottom panel, the black line is the decomposed quasar spectrum (original spectrum minus reconstructed galaxy spectrum). The brown and blue lines are the model for the continuum and \FeII\ emission; the green line is broad \hbeta\ and the cyan lines are narrow lines. The red line is the sum of all model components. The FWHM of broad \hbeta\ is shown as the black horizontal segment within the broad line profile. }
    \label{fig:fit_exmp}
\end{figure*}

\subsection{Spectral decomposition}\label{sec:decomp}

We decompose the quasar and host-galaxy spectra using the principal component analysis (PCA) method adopted by \citet{vandenberk_etal_2006} and \citet{Shen_etal_2008b}. This method assumes that the composite (quasar$+$galaxy) spectrum can be described by the combination of two independent sets of eigenspectra derived from pure galaxy and pure quasar samples. \cite{vandenberk_etal_2006} found empirically that a PCA decomposition using only the first few galaxy and quasar eigenspectra can reasonably recover the properties of the host galaxy, provided there is a significant contribution from the host galaxy in the composite spectrum. Using more eigenspectra will lead to better overall reconstruction of the full spectrum primarily around the broad lines. However, the decomposition of the quasar and host components may become ambiguous because the orthogonality between the higher-order host and quasar eigenspectra is not guaranteed.

Following \citet{vandenberk_etal_2006} and \citet{Shen_etal_2008b}, we use the first five galaxy eigenspectra and the first ten quasar eigenspectra constructed from the SDSS \citep[][]{Yip_etal_2004a,Yip_etal_2004b} to decompose our coadded quasar spectra. Fig.\ \ref{fig:fit_exmp} (top) shows an example of the spectral decomposition. A decomposition is deemed to be successful by \citet{vandenberk_etal_2006} if the host fraction $f_H$ within the wavelength range $4160\,\textrm{\AA}<\lambda<4210\,$\AA\ is greater than 0.1. This threshold of course depends on the quality of the spectrum. For our high S/N spectra, we set a threshold of $f_H=0.05$ for a successful decomposition, with additional criteria to test the reliability of the successive spectral measurements on the decomposed components. As in \citet{Shen_etal_2008b}, we obtain the pure-quasar (pure-host) spectrum by subtracting the reconstructed galaxy (quasar) spectrum from the original spectra, and use the original spectral errors for the decomposed spectra. We refer to these spectra as the (decomposed) quasar and host spectra. These spectra are used in our fitting procedures instead of the directly reconstructed spectra, because the former allows use of the associated spectral errors, and the regions with large residuals due to imperfect decomposition can be easily masked during the fitting. In addition, the reconstructed spectrum is a ``model'' rather than real data, and there could be systematic differences between the reconstructed spectrum and the real spectrum due to the limited number of eigenspectra used.

The spectral decomposition removes broad-band quasar features such as the power-law continuum and \FeII\ emission, as well as narrow-band features including quasar broad and narrow emission lines from the galaxy spectra. While the removal of the broad-band quasar feature is reasonably accurate, the removal of the quasar emission lines often shows significant residual, due to the small number of quasar eigenspectra used in the decomposition (which are insufficient to account for the full variance seen in quasar emission lines). We could add higher-order quasar eigenspectra to better reconstruct the quasar broad and narrow emission lines. But doing so would potentially over-fit the quasar component in the stellar absorption line regions and confuse the decomposition, which will impact the $\sigma_*$ measurement. Therefore we opt to use a small number of quasar eigenspectra to remove most of the broad-band quasar features, and then mask out these emission-line regions when measuring $\sigma_*$ with the decomposed host spectrum.

\begin{figure}
\centering
    \includegraphics[width=0.48\textwidth]{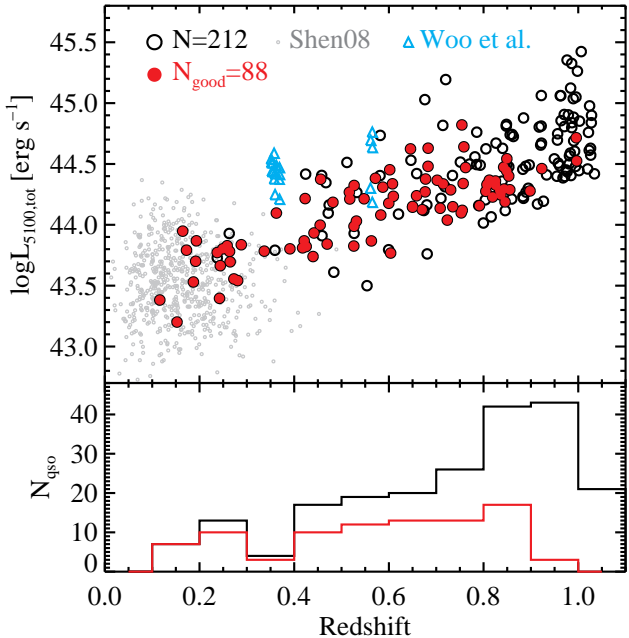}
     \caption{{\em Top:} Distribution of quasars in the luminosity-redshift plane. The continuum luminosity is estimated at restframe 5100\,\AA\ from the total spectrum. The full sample from SDSS-RM is shown in black open circles, and quasars with good $\sigma_*$ measurements are shown in red circles. The gray dots are the low-$z$ SDSS quasar sample in \citet{Shen_etal_2008b}, and the cyan open triangles are the samples at $z=0.36$ \citep{Woo_etal_2006} and at $z=0.57$ \citep[][]{Woo_etal_2008}, as we will discuss in \S\ref{sec:disc}. {\em Bottom:} Redshift histograms for the parent sample (black) and the sample with good $\sigma_*$ measurements (red). }
    \label{fig:dist}
\end{figure}

\subsection{Stellar velocity dispersion measurements}\label{sec:sigma}

The decomposed galaxy spectrum can be used to measure properties of the stellar population in the quasar host. In this work we focus on the stellar velocity dispersion, as the measurement of this quantity is relatively less affected by the imperfections in the spectral decomposition compared to other properties of the stellar population. We defer a more detailed, full analysis of the host stellar populations to future work. 

We use the {\em vdispfit} routine, publicly available in the {\em idlspec2d} product, to measure $\sigma_*$ from the decomposed galaxy spectra. This routine is used to measure stellar velocity dispersions of SDSS galaxies in the standard SDSS spectroscopic products. {\em vdispfit} is a template fitting algorithm that finds the best-fit $\sigma_*$ by minimizing the $\chi^2$ in pixel space. The templates consist of the first four PCA eigenspectra constructed from the echelle stellar spectra in the Elodie database \citep[][]{Moultaka_etal_2004} and cover restframe wavelengths from 4125\,\AA\ to 6796\,\AA. 

We restrict the fit to the wavelength range of 4125-5350\,\AA, which includes copious stellar absorption features such as the G band (4304\,\AA), the Mg Ib $\lambda\lambda$ 5167,5173,5184 triplet, and Fe (5270\,\AA), but excludes the Ca H+K $\lambda\lambda$ 3969,3934 region that is inherently unreliable for stellar velocity dispersion measurements \citep[e.g.,][]{Greene_Ho_2006a}. In addition, we mask narrow emission lines due to both galaxy and quasar excitation with a velocity window of $\pm 400\,{\rm km\,s^{-1}}$. The spectral region around the quasar broad \hbeta\ line (4760-5020\,\AA), where the PCA decomposition often shows significant residuals, is also masked for the fits. A small wavelength shift (within $\pm500\,{\rm km\,s^{-1}}$) is allowed in the fit to account for the uncertainty in the systemic redshift estimate. A fifth-order polynomial is added to account for the broad-band continuum shape. Fig.\ \ref{fig:fit_exmp} (middle) shows an example of the {\em vdispfit} fits to measure $\sigma_*$. The reported $\sigma_*$ by {\em vdispfit} is already corrected for the native SDSS instrument broadening ($\sim 65\,{\rm km\,s^{-1}}$). 

As a sanity check, we also use the direct-pixel--fitting code Penalized Pixel-Fitting (pPXF) of \citet{cappellariemsellem2004} to measure the stellar velocity dispersion. We use the stellar library from \citet{Valdes_etal_2004} with a spectral resolution of 25\,${\rm km\,s^{-1}}$ and correct the templates to the SDSS spectral resolution before fitting. Only the first two moments (radial velocity and velocity dispersion) are fitted. In addition to the radial velocity and velocity dispersion, the code solves for an eighth order multiplicative polynomial that accounts for flux calibration errors. Errors in the dispersions are derived by generating 100 mock spectra as a sum of the original spectrum and a Gaussian random realization of the spectral noise. Only the spectral range of $4000-5350$\,\AA\ is used in the fit. We mask out regions around quasar narrow but not broad emission lines. The differences in the code, spectral fitting range, and masks between the pPXF and {\em vdispfit} approaches are designed to assess the systematic uncertainty in $\sigma_*$ measurements. 

The success of the $\sigma_*$ measurements depends on a number of factors such as the subtraction of the quasar spectrum, the S/N of the resulting host galaxy spectrum, and the strength of the stellar absorption features. We define the following set of criteria for a successful $\sigma_*$ measurement: (1) the host fraction $f_H$ is greater than 0.05; (2) the median S/N per pixel of the decomposed host galaxy spectrum over restframe 4125-5350\,\AA\ is greater than 4; (3) $\sigma_*$ is measured at $>3\sigma$ statistical confidence level; and (4) visual inspection shows that the fit is not severely affected by poor PCA decomposition (i.e., significant residuals exist in the fitting region). The first three criteria remove 104 objects and the last criterion removes 20 additional objects, all with no prior knowledge of BH mass.

When compare the resulting 88 $\sigma_*$ values measured with both {\em vdispfit} and pPXF, there were 18 objects for which the difference is larger than $2\sigma$. However, 16 of these 18 objects are fits with small statistical errors in high S/N spectra, and the fractional difference between the two methods is $\lesssim 30\%$. Hence the difference between the two methods can be ascribed to the underestimation of the systematic uncertainties in the $\sigma_*$ measurements (i.e., due to differences in stellar templates, fitting ranges and emission line masks). We therefore retain the $\sigma_*$ values from the fiducial {\em vdispfit} method and indicate by the SIGMA\_WARNING flag in Table \ref{table:sample} that their measurement errors may underestimate the true uncertainty. For the remaining two objects, manual inspection suggests that the pPXF result is more robust (based on a smaller reduced $\chi^2$) and we decided to replace their {\em vdispfit} $\sigma_*$ measurements with those from the pPXF method. Our final sample of quasars with good $\sigma_*$ measurements consists of the 88 objects (which we call the ``good sample''). Fig.\ \ref{fig:dist} shows their distribution in luminosity-redshift space as compared to the parent sample. Although objects at high $z$ are preferentially lost due to our $f_H$ and S/N cuts, there is otherwise no apparent bias in the good sample compared to the parent sample in terms of object luminosity. In particular, the success rate of $\sigma_*$ measurement is quite high at $z<0.75$ ($\approx 70\%$), ensuring that our good sample is representative of the general quasar population at these redshifts. 

Although we have intentionally avoided the Ca H+K region in our $\sigma_*$ fits, these features are the strongest absorption features in most of our objects, and therefore easier to measure than those at longer wavelengths. For the sake of completeness, we also measured $\sigma_*$ in the restframe 3780--4000\,\AA\ region using pPXF (the {\em vdispfit} templates do not extend to this regime), and compile the measurements in Table \ref{table:sample}. There is broad agreement between these Ca H+K dispersions and our fiducial dispersions, but the scatter is large. We do not use these Ca H+K dispersions in our following analysis, and merely provide them in Table \ref{table:sample} for future reference. 

\begin{figure}
\centering
    \includegraphics[width=0.48\textwidth]{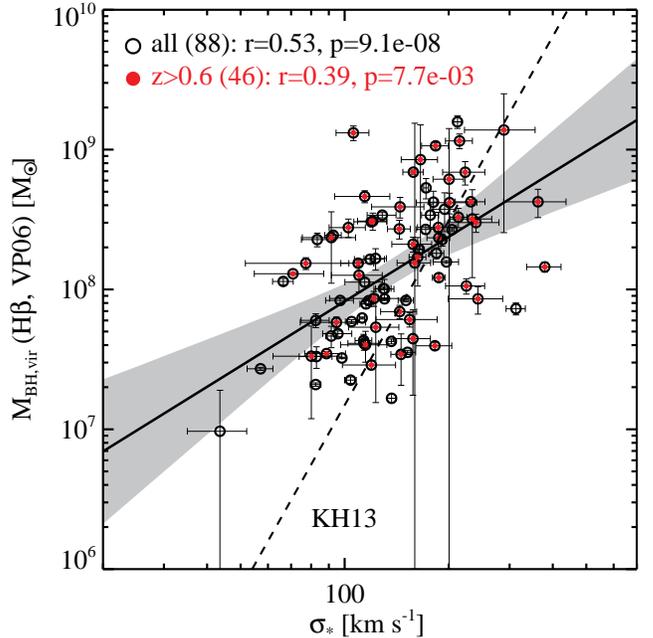}
     \caption{The relation between $\sigma_*$ and SE virial BH mass for the sample with good velocity dispersion measurements. The black points show all 88 objects, and the red points show the 46 objects at $z>0.6$. Error bars are measurement errors only, and the systematic uncertainty in the virial BH masses is $\sim 0.5$ dex \citep{Shen_2013}. Spearman's test results are shown: a significant correlation is detected both for the full sample and for objects at $z>0.6$. The black solid line is the median relation derived using the Bayesian linear regression method developed by \citet{Kelly_2007}, and the gray band indicates the 2$\sigma$ confidence range of the correlation. The vertical scatter around the best-fit relation, after correcting only for measurement errors, is 0.4\,dex. The dashed line is the relation for local inactive galaxies from \citet[][eqn.\ 7]{Kormendy_Ho_2013}. The slope of the $M_{\rm BH,fir}-\sigma_*$ relation in our quasars is significantly shallower than that of the local relation, as discussed in the text.}
    \label{fig:msigma}
\end{figure}

\begin{table*}
\caption{Catalog format}\label{table:sample}
\centering
\scalebox{1.}{
\begin{tabular}{lll}
\hline\hline
Column & Format & Description \\
\hline
RMID &      LONG                 & Index in the SDSS-RM master catalog [0-999] \\
SDSS\_NAME & STRING     & SDSS designation based on RA/DEC (hhmmss.ss$\pm$ddmmss.s) \\
RA      &     DOUBLE            & J2000 Right ascension [degree] \\
DEC  &      DOUBLE            &  J2000 Declination [degree] \\
Z       &      FLOAT           & Redshift \\ 
WAVE &    DOUBLE[4649]  &  Observed vacuum wavelength [\AA]\\
FLUX &    DOUBLE[4649] & Total flux density [${\rm 10^{-17}erg\,s^{-1}cm^{-2}\AA^{-1}}$]\\
ERR  &    DOUBLE[4649]  & Flux density error [${\rm 10^{-17}erg\,s^{-1}cm^{-2}\AA^{-1}}$]\\
FLUX\_GAL$^\dagger$ & DOUBLE[4649] & Decomposed host galaxy spectrum [${\rm 10^{-17}erg\,s^{-1}cm^{-2}\textrm{\AA}^{-1}}$]\\
FLUX\_QSO$^\dagger$ & DOUBLE[4649] & Decomposed quasar spectrum [${\rm 10^{-17}erg\,s^{-1}cm^{-2}\textrm{\AA}^{-1}}$]\\
MEDSN\_TOT  & FLOAT  & Median S/N per pixel for the full spectrum \\
MEDSN\_GAL  & FLOAT  & Median S/N per pixel for decomposed galaxy spectrum over restframe 4125-5350\,\AA \\
F\_H       & FLOAT   & Host fraction within restframe 4160--4210\,\AA; 0 if decomposition failed \\
F\_H\_5100  & FLOAT &  Host fraction at restframe 5100\,\AA\\
SIGMA   &  DOUBLE    & Fiducial stellar velocity dispersion [km/s]; 0 if not available \\
SIGMA\_ERR & DOUBLE & Measurement error in SIGMA; $-1$ if not available \\
SIGMA\_ERR\_WARNING & LONG  & 1 if SIGMA\_ERR may underestimate the systematic uncertainty\\
SIGMA\_HK  &  DOUBLE    & Stellar velocity dispersion based on Ca H+K [km/s]; 0 if not available \\
SIGMA\_HK\_ERR & DOUBLE & Measurement error in SIGMA\_HK; $-1$ if not available \\
LOGL5100\_TOT &  DOUBLE & Total continuum luminosity at restframe 5100\,\AA\ [erg/s]; 0 if not covered \\
LOGL5100\_TOT\_ERR & DOUBLE & Measurement error in LOGL5100\_TOT \\
LOGL5100\_QSO &  DOUBLE & Quasar continuum luminosity at restframe 5100\,\AA\ [erg/s]; 0 if not covered \\
LOGL5100\_QSO\_ERR & DOUBLE & Measurement error in LOGL5100\_QSO \\
FWHM\_HB      & DOUBLE    &  FWHM of broad \hbeta\ [km/s] \\
FWHM\_HB\_ERR  & DOUBLE  & Measurement error in FWHM\_HB \\
LOGMBH\_VP06  & DOUBLE  &  SE virial mass estimate based on the VP06 \hbeta\ recipe [$M_\odot$] \\
LOGMBH\_VP06\_ERR  & DOUBLE  & Measurement error in LOGMBH\_VP06 \\
\hline\\
\end{tabular}
}
\begin{tablenotes}
      \small
      \item NOTE. --- Format of the fits catalog containing the data and measurements of the 212 quasars used in this paper, ordered by RMID. The 88 quasars with good $\sigma_*$ measurements are identified with ${\rm SIGMA}>0$. The SIGMA\_ERR\_WARNING flag is set if the difference between the fiducial $\sigma_*$ measurement and that from pPXF is larger than 2$\sigma$. The errors reported in SIGMA\_ERR may be an underestimation in these cases. The $\sigma_*$ measurements based on Ca H$+$K features (SIGMA\_HK and SIGMA\_HK\_ERR) are not suitable for studying the $M_\bullet-\sigma_*$ relation, and are only included here for future reference. The full catalog is available in the electronic version of the paper.  \\
      $^\dagger$The decomposed galaxy (quasar) spectra are obtained by subtracting the reconstructed quasar (galaxy) model from the full spectrum, and only cover the rest-frame 3450--7498\,\AA\ range, i.e., the overlapping wavelength range in the quasar and galaxy PCA eigenspectra from \citet{Yip_etal_2004a} and \citet{Yip_etal_2004b}.
      \end{tablenotes}
\end{table*}

\subsection{Single-epoch virial mass estimates}\label{sec:bh}

To apply single-epoch virial BH mass estimators for our quasars, we follow earlier work \citep[e.g.,][]{Shen_etal_2008a,Shen_etal_2011} to fit the quasar spectrum with continuum emission, \FeII\ emission, and broad and narrow emission lines around the \hbeta\ region. Fig.~\ref{fig:fit_exmp} (bottom) shows an example of the fits to the quasar spectrum. The continuum luminosity (i.e., $L_{5100}\equiv \lambda L_{\lambda,5100\,\textrm{\AA}}$) and broad line width are measured from the best-fit model. We use the Monte Carlo approach described in \citet[][]{Shen_etal_2011} to estimate measurement errors for these quantities. We choose the FWHM to characterize the width of the broad \hbeta\ line instead of using the line dispersion $\sigma_{\rm line}$ for the reasons discussed in \citet[][and references therein]{Shen_2013}. However, we will experiment with different virial BH mass recipes that account for the difference between using FWHM and $\sigma_{\rm line}$. 

The standard SE virial BH mass estimator based on \hbeta\ is: 
\begin{equation}
\log \left(\frac{M_{\rm BH,vir}}{M_\odot}\right)=a + b\log \left(\frac{L_{5100}}{{\rm 10^{44}\,erg\,s^{-1}}}\right)+c\log \left(\frac{{\rm FWHM}}{\rm km\,s^{-1}}\right)\ ,
\end{equation}
where coefficients $a$, $b$ and $c$ are determined empirically using the local RM AGN sample in different recipes \citep[e.g.,][]{Vestergaard_Peterson_2006,Wang_etal_2009,Feng_etal_2014}. The luminosity range in the local RM AGN sample encompasses most of our quasars. 

Our fiducial mass recipe is that of \citet[][eqn.\ 5; hereafter VP06]{Vestergaard_Peterson_2006} with $a=0.91$, $b=0.5$ and $c=2$. In VP06, the continuum luminosity $L_{5100}$ was not corrected for host starlight contamination, hence we use the total 5100\,\AA\ luminosity in this recipe.\footnote{We also tried using the quasar-only 5100\,\AA\ luminosity in the VP06 recipe and found similar conclusions. The systematic uncertainties in the virial BH mass estimates (including the zero-point uncertainty) are much larger than this detail \citep[e.g.,][]{Shen_2013}.} We use the VP06 virial BH masses in our following analysis. 

We also consider an alternative recipe in \citet[][eqn.\ 14; hereafter F14]{Feng_etal_2014} with $a=3.602$, $b=0.504$ and $c=1.2$. The F14 recipe requires the host-corrected quasar $L_{\rm 5100,qso}$, which is measured from the decomposed quasar spectrum. This recipe relaxes the assumption that BH mass is proportional to the square of FWHM, and uses the best-fit slope on FWHM from the calibration against the local sample of AGN with RM masses \citep[also see, e.g.,][]{Wang_etal_2009}. Since these RM masses were computed using the line dispersion $\sigma_{\rm line}$ in the rms spectrum rather than FWHM, this recipe effectively accounts for the nonlinear relation between FWHM and $\sigma_{\rm line}$ \citep[e.g.,][]{Collin_etal_2006,Shen_Ho_2014}. We use this alternative set of BH mass estimates to demonstrate the ambiguities in measuring BH masses for high-$z$ quasars with the SE methods, as discussed in \S\ref{sec:disc}.

The measurement errors in $\log M_{\rm BH,vir}$ are propagated from the measurement errors in FWHM and $\log L_{5100}$, but the dominant uncertainty in $\log M_{\rm BH,vir}$ is the systematic uncertainty of single-epoch virial masses, which can be as large as $\sim 0.5$ dex \citep[e.g.,][]{Shen_2013}. We successfully derived $M_{\rm BH,vir}$ estimates for all quasars in our parent sample except for a few cases where restframe 5100\,\AA\ is not covered by the spectrum. None of these sources were included in the good sample. 

The VP06 SE mass recipe uses a constant average virial coefficient that normalizes the virial BH masses of nearby reverberation mapping AGN to follow the local $M_\bullet-\sigma_*$ relation in inactive galaxies \citep[e.g.,][]{Onken_etal_2004}. An implicit assumption in our investigations is that this average virial coefficient does not evolve with redshift, which is a reasonable assumption since this is largely a geometrical factor to account for the structure of the quasar broad-line region \citep[e.g.,][]{Shen_2013}.

\section{Results}\label{sec:result}

\begin{figure*}
\centering
    \includegraphics[width=0.98\textwidth]{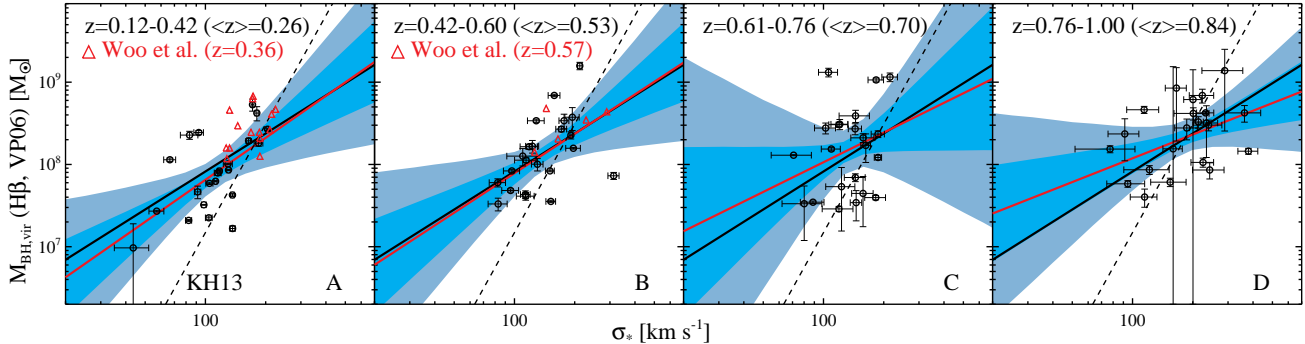}
     \caption{Redshift evolution of the $M_{\rm BH,vir}-\sigma_*$ relation. The full sample (88 objects) is divided into four equal-number subsamples by redshift. For each subsample, we perform the same Bayesian linear regression on the data.  In each redshift bin, the red line is the median relation from the Bayesian fit; the cyan and light blue shaded regions are the $1\sigma$ and $2\sigma$ confidence ranges of the fit. For reference, the median relation for the full sample is shown as the black solid line, and the local relation in \citet{Kormendy_Ho_2013} is shown as the black dashed line. We do not detect significant differences in the $M_{\rm BH,vir}-\sigma_*$ relation from the lowest redshift bin to the highest redshift bin in our sample. However, there is an indication that the relation becomes flatter towards higher redshift, which is further discussed in \S\ref{sec:disc}. We also show the samples in \citet[][$z=0.36$]{Woo_etal_2006} and \citet[][$z=0.57$]{Woo_etal_2008} in red open triangles. The objects in Woo et~al.\ are more luminous than our quasars in the relevant redshift bins (see Fig.\ \ref{fig:dist}). Since Woo et~al.\ did not report FWHM measurements, we took BH mass estimates directly from their papers.}
    \label{fig:msigma_evo}
\end{figure*}

\begin{table}
\caption{Linear Regression Results}\label{table:reg}
\centering
\scalebox{1.}{
\begin{tabular}{lcccc}
\hline\hline
Sample & $\bracket{z}$ & $\alpha$ & $\beta$ & Intrinsic scatter \\
\hline\\
full                     & 0.60 & $8.377_{-0.066}^{+0.067}$ & $1.535_{-0.303}^{+0.304}$ & $0.406_{-0.031}^{+0.035}$ \\
$z>0.6$             & 0.76 & $8.395_{-0.088}^{+0.088}$ & $1.081_{-0.492}^{+0.511}$ & $0.440_{-0.048}^{+0.059}$ \\
$0.12<z<0.42$  & 0.26 & $8.324_{-0.180}^{+0.170}$ & $1.695_{-0.604}^{+0.614}$ & $0.410_{-0.063}^{+0.082}$ \\
$0.42<z<0.60$  & 0.53 & $8.372_{-0.138}^{+0.137}$ & $1.592_{-0.660}^{+0.677}$ & $0.414_{-0.062}^{+0.082}$ \\
$0.61<z<0.76$  & 0.70 & $8.388_{-0.235}^{+0.233}$ & $1.199_{-1.216}^{+1.231}$ & $0.536_{-0.083}^{+0.109}$ \\
$0.76<z<1.00$  & 0.84 & $8.364_{-0.091}^{+0.097}$ & $0.960_{-0.588}^{+0.598}$ & $0.381_{-0.068}^{+0.091}$ \\\\
\hline\\
\end{tabular}
}
\begin{tablenotes}
      \small
      \item NOTE. --- Linear regression of the relation $\log (M_{\rm BH,vir}/M_\odot)=\alpha + \beta\log(\sigma_*/{\rm 200\,km\,s^{-1}})$ on the data, using the Bayesian approach described in \citet{Kelly_2007}. The intrinsic scatter refers to the scatter in $\log M_{\rm BH}$ around the best-fit relation, after only accounting for measurement errors. Systematic uncertainties in $\log M_{\rm BH,vir}$ are included in this intrinsic scatter. The 1$\sigma$ errors are estimated from the 16\% and 84\% percentiles of the posterior distribution of the parameter from the Bayesian fits. 
\end{tablenotes}
\end{table}

\begin{figure}
\centering
    \includegraphics[width=0.48\textwidth]{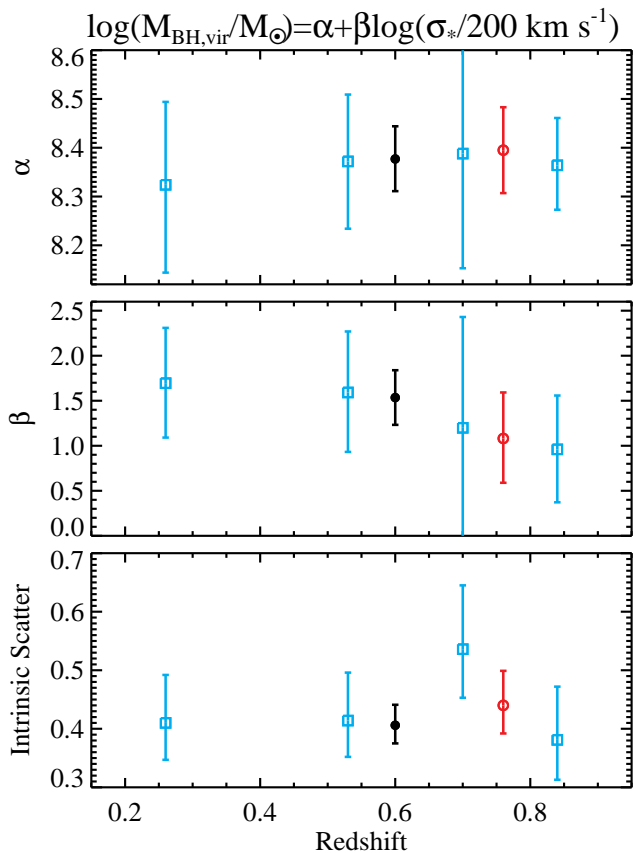}
     \caption{Redshift evolution of the best-fit parameters in the $M_{\rm BH,vir}-\sigma_*$ relation defined by Eqn.\ (\ref{eqn:msigma}) and tabulated in Table \ref{table:reg}. The black filled circles are for the full sample, the red open circles are for the $z>0.6$ subsample, and the cyan open squares are for the four redshift bins.}
    \label{fig:para_evo}
\end{figure}

We show our $\sigma_*$ and $\log M_{\rm BH,vir}$ measurements in Fig.\ \ref{fig:msigma}. The full sample is shown in black circles with measurement errors only. The Spearman test finds a strong probability of a correlation, as shown by the values presented in Fig.\ \ref{fig:msigma}. For the first time, we were able to detect directly a correlation between $\sigma_*$ and virial BH mass estimates at $\bracket{z}=0.6$, while earlier studies did not have the statistics and dynamic range to clearly detect a correlation at $z>0.3$ \citep[e.g.,][]{Woo_etal_2006,Woo_etal_2008}. A moderately strong correlation is still present if we restrict the analysis to the subset of 46 $z>0.6$ objects (red circles). The correlation for the high-$z$ subset is similar to the relation for the full sample, suggesting negligible evolution, as further demonstrated below.  

We use the Bayesian linear regression method in \citet{Kelly_2007} to fit for $\log M_{\rm BH,vir}$ as a function of $\log\sigma_*$: 
\begin{equation}\label{eqn:msigma}
\log \left(\frac{M_{\rm BH,vir}}{M_\odot}\right)=\alpha + \beta\log\left(\frac{\sigma_*}{{\rm 200\ km\,s^{-1}}}\right)\ ,
\end{equation} 
where $\alpha$ and $\beta$ are the intercept (at $\sigma_*=200\ {\rm km\,s^{-1}}$) and slope of the regression, and we use $\epsilon$ to denote the intrinsic scatter in $\log M_{\rm BH}$ around the best fit. This Bayesian regression method properly accounts for the intrinsic scatter, thus the best-fit slope is not biased due to a few points with small measurement errors \citep[e.g., see discussion in][]{Tremaine_etal_2002}. The results are summarized in Table \ref{table:reg}, where the significantly non-zero (at 5$\sigma$) slope $\beta$ for the full sample again suggests a strong correlation. We only consider the measurement errors of $\log M_{\rm BH,vir}$ in the fit, and thus the intrinsic scatter $\epsilon$ includes the contribution from systematic errors in $\log M_{\rm BH,vir}$. We also tried adding 0.5\,dex systematic errors on $\log M_{\rm BH,vir}$, or using alternative regression methods \citep[e.g.,][]{Tremaine_etal_2002}, but did not find significantly different results for the intercept and slope. 

In Fig.\ \ref{fig:msigma}, the solid line represents the relation with the median intercept $\alpha=8.38$ and slope $\beta=1.54$ in the posterior distribution from the Bayesian fit, and the gray band indicates the 2$\sigma$ confidence range of the correlation. The intrinsic scatter is estimated to be 0.4\, dex. Although there is a strong correlation between $M_{\rm BH,vir}$ and $\sigma_*$ in our sample, the best-fit slope is significantly shallower than that of the $M_{\bullet}-\sigma_*$ relation for local inactive galaxies \citep[][eqn.\ 7, with $\beta=4.38$]{Kormendy_Ho_2013}, which is shown as the dashed line. We discuss this discrepancy in detail in \S\ref{sec:disc}. On the other hand, the average BH mass at $\sigma_*=200\ {\rm km\,s^{-1}}$ is $\sim 2.5\times 10^8\ M_\odot$, consistent with the local relation. 

The large dynamic range in luminosity and redshift of our sample, as well as the homogeneous analysis, make it possible to investigate potential redshift evolution of the $M_{\rm BH,vir}-\sigma_*$ relation within our sample. We divide the full 88 objects into four equal-number subsamples by redshift, and show the results in Fig.\ \ref{fig:msigma_evo}. We perform linear regression with the Bayesian method to the data in each subsample, and show the median relation (red lines) and the 1$\sigma$ (2$\sigma$) confidence range in cyan (light blue) shaded bands in Fig.\ \ref{fig:msigma_evo}. A correlation is inferred (albeit at low statistical significance) even in the highest redshift bin with $\bracket{z}=0.84$. The regression results are summarized in Table \ref{table:reg} and plotted in Fig.\ \ref{fig:para_evo}.

The median relation for the full sample is shown in black lines, and is consistent within 1$\sigma$ with the relations at different redshifts. Therefore we do not detect any redshift evolution in the $M_{\rm BH,vir}-\sigma_*$ relation. In particular, the mean BH mass at $\sigma_*=200\ {\rm km\,s^{-1}}$ is remarkably stable, and is consistent with the value from the local relation in \citet{Kormendy_Ho_2013}. There is, however, an indication that the relation flattens towards higher redshift (see Table \ref{table:reg}), as further discussed in \S\ref{sec:disc}. 

\section{Discussion}\label{sec:disc}

\begin{figure}
\centering
    \includegraphics[width=0.48\textwidth]{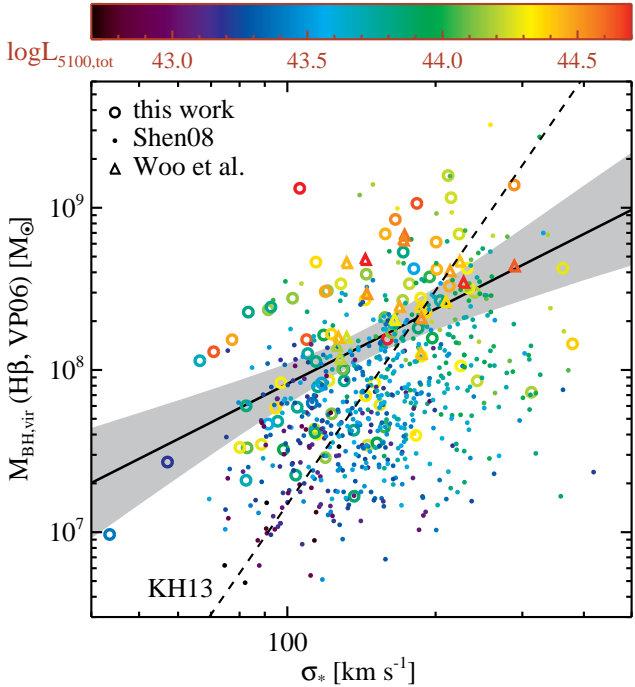}
     \caption{Comparison between our results and earlier results. All data points are color-coded by the total continuum luminosity $L_{5100}$. Measurement errors have been suppressed for clarity. The \citet{Shen_etal_2008b} sample includes quasars at lower redshift, and their luminosities are substantially lower than our quasars, as indicated in Fig.\ \ref{fig:dist}. The \citet{Shen_etal_2008b} sample shows systematic offset towards lower BH masses, which is an indication that our high-$z$ sample is missing low-mass BHs at fixed $\sigma_*$ due to the luminosity threshold. The samples in \citet{Woo_etal_2006} and \citet{Woo_etal_2008} are shown in open triangles. They overlap with our quasars with comparable luminosities, but are limited to a narrow dynamic range in $\sigma_*$ and BH mass. Since Woo et~al.\ did not report FWHM measurements, we took BH mass estimates directly from their papers.
     The black line and gray shaded band are the Bayesian linear regression results as in Fig.\ \ref{fig:msigma}, and the dashed line denotes the local relation from \citet{Kormendy_Ho_2013}.}
    \label{fig:msigma_comp}
\end{figure}

The results presented above should be interpreted with great caution. Below we discuss a list of potential caveats that went into the above analyses and interpretation, and our resolutions. \\


{\em Spectral decomposition}\\
 
Measuring the host stellar velocity dispersion in quasars is always challenging because the stellar absorption features are diluted by the significant contribution from the quasar. The PCA decomposition technique attempts to subtract the quasar continuum and broad line emission (including the optical \FeII\ complex that extends over a broad spectral range). However, it is difficult to ensure a clean subtraction of the quasar contribution, and this may affect the dispersion measurements. To mitigate the residual impact of quasar emission, we masked prominent broad and narrow emission line regions when fitting  the decomposed galaxy spectrum for $\sigma_*$; while broad-band quasar residuals are accounted for with low-order polynomials in the $\sigma_*$ fit. We believe in general this approach should work well.\\

{\em $\sigma_*$ measurements} \\

The main hurdle in our $\sigma_*$ measurement is the S/N of the stellar absorption features in the decomposed galaxy spectrum. We have set a threshold of median ${\rm S/N}>4$ per pixel to ensure a reasonable $\sigma_*$ measurement (the median S/N per pixel in the galaxy spectrum is $\sim 15$ for our good sample). We have also tested the sensitivity of the $\sigma_*$ measurement to changes in the stellar templates, fitting code, spectral fitting range and masks, and found our fiducial $\sigma_*$ measurements are robust against these details. The ultimate limitation on $\sigma_*$ measurements in quasars likely lies in the systematics from spectral decomposition in extremely quasar-dominated spectra, spectral calibration and template matching -- our measurements are far from reaching that limit. For example, it is still difficult to measure $\sigma_*$ in the most luminous PG quasars at low $z$ (which are highly biased relative to the general quasar population in terms of luminosity and radio-loudness) with a huge quasar-host contrast -- S/N is not the limiting factor there\footnote{A promising approach for measuring $\sigma_*$ in these most luminous quasars is to use near-infrared stellar absorption features assisted with adaptive optics to minimize the impact of the quasar continuum \citep[e.g.,][]{Dasyra_etal_2007,Watson_etal_2008,Grier_etal_2013}. But even in this case, the measurement of $\sigma_*$ can be extremely difficult in the most luminous quasars.}. 

The completeness in $\sigma_*$ measurements for quasar samples poses another concern. If a significant fraction of objects with high quasar-to-host contrast are preferentially lost from the sample due to the failure of $\sigma_*$ measurements, there is a potential selection bias in the resulting $M_{\rm BH,vir}-\sigma_*$ relation, such that the average $\sigma_*$ at fixed BH mass may be biased high. Fortunately for our sample, the completeness in $\sigma_*$ measurements is high ($\approx 70\%$) at $z<0.75$, meaning that the potential bias due to sample incompleteness is not a severe issue for the bulk of our sample. However, it may become important for the highest redshift bin shown in Fig.\ \ref{fig:msigma_evo}, where the fraction of quasars with measured $\sigma_*$ is low mainly due to small host fractions. There is no easy way to quantify this effect -- deeper spectroscopy is required to increase the chance of measuring $\sigma_*$ in these objects.

The next question is what $\sigma_*$ we are actually measuring. SDSS-BOSS spectra collect light enclosed in a 2\arcsec\ diameter circle on the sky, which translates to a physical radius of $3.3-7.8\,$kpc at $0.2<z<0.9$. Therefore our $\sigma_*$ is measured from the integrated light from the bulge and a potential disk component at high $z$. We are ignorant about the morphological types of our quasars and their redshift dependence, so an object-by-object correction to the bulge $\sigma_*$ is difficult. We have tried the empirical correction for aperture effects in \citet{Shen_etal_2008b}, and did not find significant changes in our results. While we caution that some of the scatter in our results may arise from the failure to resolve the host light into disk and bulge components, the estimated bias in $\sigma_*$ due to the inclusion of a rotating disk is $\lesssim 16\%$ based on local AGN samples \citep[e.g.,][]{Bennert_etal_2014}. Thus this effect is negligible given the measurement uncertainties and intrinsic scatter in Fig.\ \ref{fig:msigma}.\\

{\em BH mass estimates} \\

A larger concern is in the BH mass estimates based on the SE methods. As discussed at length in \citet{Shen_2013}, there are currently large systematic uncertainties in the SE virial masses for high-$z$ quasars at the $\sim 0.5\,$ dex level (both in scatter and in zeropoint), and systematic differences between different recipes. 

To investigate the difference due to alternative SE mass recipes, we have repeated the analysis based on the F14 mass recipe for \hbeta. We still find a significant correlation between $M_{\rm BH,vir}$ and $\sigma_*$ in different redshift bins, but the slope of the correlation is even shallower than that based on the VP06 masses. This exercise highlights the ambiguity in estimating BH masses using the SE method, and the difficulty in comparing different studies using different BH mass recipes. Nevertheless, the small samples at $z=0.36$ and $z=0.57$ in \citet{Woo_etal_2006} and \citet{Woo_etal_2008}, when matched in luminosity and redshift to our quasars, fall consistently within our data points in Fig.\ \ref{fig:msigma_evo}. 

Since all SE mass recipes explicitly depend on luminosity, statistical biases will arise from samples with a threshold luminosity (as we further discuss below).\\

{\em Statistical biases} \\

There are several statistical biases due to quasar-based sample selection and BH mass uncertainties that have been previously discussed \citep[e.g.,][]{Adelberger_Steidel_2005,Salviander_etal_2007,Lauer_etal_2007b,Shen_etal_2008a,Shen_Kelly_2010,Schulze_Wisotzki_2011}. Below we discuss these effects, and provide qualitative assessments of their implications on the interpretation of Fig.\ \ref{fig:msigma} and Fig.\ \ref{fig:msigma_evo}. 

\citet{Lauer_etal_2007b} pointed out that \citep[also see][]{Adelberger_Steidel_2005,Salviander_etal_2007}, because there is intrinsic scatter in the $M_\bullet-\sigma_*$ relation, samples selected based on a threshold in quasar luminosity will preferentially select overmassive BHs with respect to galaxy stellar velocity dispersion, assuming that more luminous quasars represent more massive BHs on average. \citet{Shen_Kelly_2010} further suggested that \citep[based on the earlier suggestion in][]{Shen_etal_2008a}, because SE virial mass estimates explicitly depend on luminosity and there is non-negligible random scatter between instantaneous luminosity and line width (i.e., the origin of the uncertainties in SE virial masses), quasar samples with a threshold luminosity are, on average, biased high in their SE virial mass estimates.\footnote{A detailed derivation of this virial BH mass bias is presented in Sec 3.3 of \citet[][]{Shen_2013}, which is referred to as the luminosity-dependent bias in SE virial BH masses.} These two statistical biases will work independently to produce a BH mass offset relative to $\sigma_*$ in high-$z$, hence high-luminosity, quasar samples; this BH mass offset is more prominent at lower $\sigma_*$ (hence flattening the slope in $M_\bullet-\sigma_*$), where under-massive BHs (relative to $\sigma_*$) are more easily lost due to the sample luminosity threshold. In addition, both effects become worse at higher threshold luminosities, which qualitatively explains the flattening in the slope of the $\log M_{\rm BH,vir}-\sigma_*$ relation with redshift seen in Fig.\ \ref{fig:msigma_evo}. A corollary of this result is that if we probe sufficiently faint quasar luminosities at high $z$, we should recover the population of undermassive BHs at fixed $\sigma_*$.

To strengthen these arguments, we perform a test using the low-$z$ quasar sample in \citet{Shen_etal_2008b}. Under the premise that there is no redshift evolution in the intrinsic $M_\bullet-\sigma_*$ relation, we plot the sample in \citet[][hereafter the Shen08 sample]{Shen_etal_2008b}, and compare with our data in Fig.\ \ref{fig:msigma_comp}. The Shen08 sample is a low-$z$ SDSS quasar sample, analyzed using the same PCA method to decompose the spectrum, and measure host stellar velocity dispersion, quasar continuum luminosity and broad \hbeta\ FWHM. To be consistent with our analysis, we use the total continuum luminosity $L_{5100}$ from their compiled measurements, and scale their FWHM values upwards by a constant 0.05\,dex to match our own measurements for the objects in common with the catalog of \citet{Shen_etal_2011}. We then use the VP06 recipe to compute virial BH masses for their sample. The Shen08 sample is shown in Fig.\ \ref{fig:msigma_comp} with small circles. This low-$z$ sample has significantly lower luminosities than our quasars (see Fig.\ \ref{fig:dist}), and as expected, they are systematically offset from our quasars at fixed $\sigma_*$. In fact, a systematic trend of BH mass offset as luminosity increases is present in the Shen08 sample as well. These results are consistent with the above argument that high-$z$ luminosity-threshold samples suffer from statistical biases in their average BH mass relative to the expectation from the mean $M_\bullet-\sigma_*$ relation. 

\begin{figure}
\centering
    \includegraphics[width=0.48\textwidth]{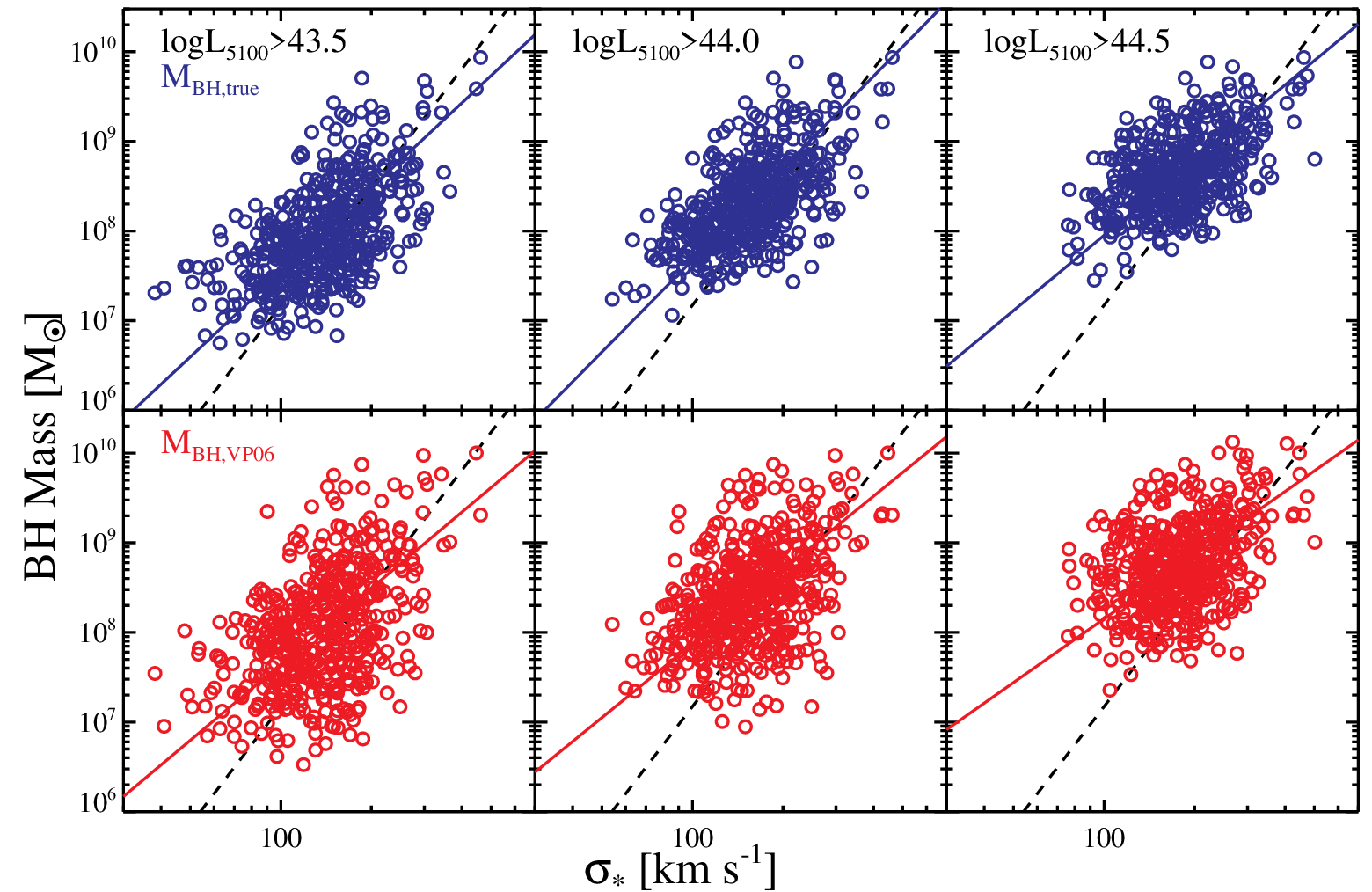}
     \caption{Illustration of the statistical biases discussed in \S\ref{sec:disc}. We simulate a mock distribution of true BH masses using the observed local stellar velocity distribution function \citep{Bernardi_etal_2010}, convolved with the local $M_\bullet-\sigma_*$ relation \citep{Kormendy_Ho_2013} with a scatter of 0.44 dex in $M_\bullet$ \citep[e.g.,][]{Gultekin_etal_2009}. We then assign quasar luminosities assuming a lognormal Eddington ratio ($\lambda$) distribution with a mean of $\langle\log\lambda\rangle=-1$ and a scatter of 0.3 dex. We then generate FWHMs and virial BH masses following the approach detailed in \citet{Shen_2013}. The distribution of FWHM is lognormal with a scatter of 0.18 dex, and a mean depending on the true BH mass and the mean luminosity at fixed true mass. By design, the simulated virial BH masses have an uncertainty of 0.39 dex around their true masses, roughly consistent with current estimates \citep[e.g.,][]{Shen_2013}. We then impose different threshold luminosities to select a random set of 500 quasars from the mock sample, shown in open circles. The top panels show the distribution in terms of true BH mass, and the bottom panels show the corresponding luminosity-threshold mock samples in terms of virial BH masses. The solid lines represent the best-fit linear regression results using the \citet{Kelly_2007} method. The black dashed lines denote the local relation from \citet{Kormendy_Ho_2013}. As the luminosity threshold increases, we preferentially select over-massive BHs at the low-$\sigma_*$ end, which causes a flattening in the regression slope. The top panels demonstrate the \citet{Lauer_etal_2007b} bias, while the bottom panels demonstrate the combined effect of the \citet{Lauer_etal_2007b} bias and the \citet{Shen_Kelly_2010} bias. }
    \label{fig:sim_bias}
\end{figure}

To further illustrate the effects of these two statistical biases, we perform simple simulations of luminosity-threshold quasar samples. We start by simulating a distribution of quasars with known velocity dispersions and true BH masses using the local stellar velocity dispersion function \citep{Bernardi_etal_2010} convolved with the local $M_\bullet-\sigma_*$ relation \citep{Kormendy_Ho_2013}. We assume an intrinsic scatter of 0.44 dex in $M_\bullet$ at fixed $\sigma_*$ for all morphological types of galaxies \citep[e.g.,][]{Gultekin_etal_2009}. With the true BH masses, we generate quasar bolometric luminosity $L_{\rm bol}$ using a lognormal distribution of Eddington ratio, $\lambda\equiv L_{\rm bol}/L_{\rm Edd}$, where $L_{\rm Edd}=1.26\times 10^{38}(M_\bullet/M_\odot)\ {\rm erg\,s^{-1}}$ is the Eddington luminosity. We choose a mean Eddington ratio $\langle\log\lambda\rangle=-1$ and a scatter $\sigma_L=0.3$ dex \citep[e.g.,][]{Shen_etal_2008a,Shen_Kelly_2012}. We assume $L_{\rm bol}=10L_{5100}$ to get $L_{5100}$. We then follow \citet{Shen_2013} to generate FWHM and virial BH masses using the VP06 recipe. The FWHMs are generated assuming a lognormal distribution, with the mean depending on true BH mass and mean luminosity at fixed true mass, and a scatter of $\sigma_{\rm FWHM}=0.18$ dex. This approach ensures that the virial BH masses are on average unbiased (when no luminosity constraints are imposed), and that the FWHM distribution of quasars is consistent with observation \citep[e.g.,][]{Shen_etal_2008a}. We assume no correlation between the luminosity dispersion $\sigma_L$ and FWHM dispersion $\sigma_{\rm FWHM}$ at fixed true BH mass. Therefore the uncertainty of the virial masses is $\sigma_{\rm vir}=\sqrt{(0.5\sigma_L)^2 + (2\sigma_{\rm FWHM})^2}=0.39$ dex, roughly consistent with expectation \citep[e.g.,][]{Shen_2013}. In addition, this approach yields a maximum luminosity-dependent bias in the virial BH masses \citep[e.g.,][]{Shen_etal_2008a,Shen_Kelly_2010,Shen_2013}. Again, we refer the reader to \citet[][Sec 3.3]{Shen_2013} for a more elaborate derivation of this virial BH mass bias. 

The results are shown in Fig.\ \ref{fig:sim_bias}, where we select a random set of 500 quasars from the simulated sample by imposing different luminosity thresholds. We have shuffled the $\sigma_*$ values by 0.1 dex to mimic measurement errors. The top panels show the simulated distributions in terms of true BH masses, which demonstrate the effect of the \citet{Lauer_etal_2007b} bias. The bottom panels show the simulated distributions in terms of virial BH masses, which demonstrate the combined effect of the \citet{Lauer_etal_2007b} bias and the virial BH mass bias. In both cases, the luminosity-threshold samples preferentially select overly-massive BHs at the low-$\sigma_*$ end, and the best-fit regression slope (the solid lines) is significantly shallower than the intrinsic slope (the dashed lines). This simple simulation qualitatively explains the observed trends in our data, and provides strong evidence that statistical biases are at work. 


\citet{Schulze_Wisotzki_2011} suggested a scenario where the active fraction of quasars may be a function of BH mass. Coupled with the intrinsic scatter in the $M_\bullet-\sigma_*$ relation, this effect would also produce deviations from the mean $M_\bullet-\sigma_*$ relation in samples with quasar-based selection. However, at low $z$, there appears to be no significant difference in the $M_\bullet-\sigma_*$ relation in inactive and active galaxies \citep[e.g.,][]{Woo_etal_2013,Bennert_etal_2014}. The situation, however, may be different at high $z$. 


The ``intrinsic'' scatter in BH mass around the best-fit regression listed in Table \ref{table:reg} is the combination of the intrinsic scatter in the $M_\bullet-\sigma_*$ relation and the scatter in SE virial BH mass estimates around true BH masses. At face value, this suggests that the systematic uncertainty in SE virial BH masses is less than $\sim 0.4\,$dex, since there must be intrinsic scatter in the $M_\bullet-\sigma_*$ relation. This statement is incorrect, however: because of the luminosity threshold and the way the two biases operate, we are not sampling the full distribution of objects and thus do not probe the true uncertainty in SE virial masses. If we were to select a sample with fixed continuum luminosity and fixed line width, then the dispersion in the sample virial BH masses would be zero, while the uncertainty in these BH mass estimates would still be $\sim 0.5\,$dex. \citet[][Sec 3.3]{Shen_2013} describes in detail how to distinguish these different concepts. In brief, the observed distribution in the SE virial masses for a given sample is the convolution of the underlying BH mass distribution with the luminosity selection function and the SE mass error distribution. Luminosity-threshold quasar samples often show a dispersion of virial masses that can be smaller than the nominal mass uncertainty \citep[e.g.,][]{Shen_etal_2008a,Shen_Kelly_2012}. These two are equal only when: (1) the true masses have zero sample dispersion; and (2) the error in SE virial masses does not depend on luminosity -- neither of these are true, as discussed in \citet{Shen_2013}. 

Similarly, even if we can account for the scatter in SE BH masses around true BH masses, the dispersion in true BH mass for the luminosity threshold sample is not the intrinsic scatter around the mean $M_\bullet-\sigma_*$ relation, precisely due to the Lauer et~al.\ bias: we are sampling the upper envelope of the $M_\bullet-\sigma_*$ relation rather than the full distribution. \citet{Jahnke_Maccio_2011} proposed an interesting scenario, based on an earlier suggestion by \citet{Peng_2007}, that the $M_\bullet-\sigma_*$ relation can be naturally produced by the averaging process during mergers of galaxies with initially uncorrelated BH mass and galaxy properties. Such a scenario predicts an increasing intrinsic scatter in the $M_\bullet-\sigma_*$ relation with redshift, and is testable with observations. Unfortunately, we cannot test this prediction with our data without a more careful forward modeling of the complications just described. Nevertheless, the nearly constant scatter in Table \ref{table:reg}, albeit only measuring the upper envelop of the $M_\bullet-\sigma_*$ relation due to selection effects, seems to favor a non-evolving intrinsic scatter in the $M_\bullet-\sigma_*$ relation.  

Our observations are consistent with \citet{Woo_etal_2006} and \citet{Woo_etal_2008}: if we look only at their samples at $z=0.36$ and $z=0.57$, which are at comparable luminosities (see Fig.\ \ref{fig:dist}), then there is little evolution. A strong evolution was inferred only when they compared to a low-$z$ and low-luminosity sample, as shown here as well (Fig.\ \ref{fig:msigma_comp}). Given the limited dynamic range in luminosity in the samples of \citet{Woo_etal_2006} and \citet{Woo_etal_2008}, this evolution manifested as an offset from the local relation. However, the low-$z$ samples have on average much lower luminosity than the high-$z$ samples, so this comparison suffers from the statistical biases discussed above. The \citet{Woo_etal_2006,Woo_etal_2008} samples are on average more luminous than our quasars at the relevant redshifts (see Fig.\ \ref{fig:dist}), and they do tend to lie in the upper envelop of the distribution of our quasars in the $M_{\rm BH,vir}-\sigma_*$ plane, as shown in Fig.\ \ref{fig:msigma_evo}. By going to lower luminosities at these redshifts, our sample was able to recover some (but not all) of the low-mass BHs at fixed $\sigma_*$ that were missed in earlier samples. All these observations are consistent with the expectation from statistical biases that we demonstrated in Fig.\ \ref{fig:sim_bias} with simple models.  

Our interpretation of the observed $M_{\rm BH,vir}-\sigma_*$ relation at high $z$ is therefore fundamentally different from Woo et~al., in that we believe the differences seen between high-$z$ and low-$z$ samples are mainly not due to real evolution, but largely caused by selection biases and BH mass uncertainties in luminosity-threshold quasar samples, as emphasized in several recent studies \citep[e.g.,][]{Salviander_etal_2007,Lauer_etal_2007b,Jahnke_etal_2009,Shen_Kelly_2010,Schulze_Wisotzki_2011,Shen_2013,Salviander_Shields_2013,Schramm_Silverman_13, Matsuoka_etal_2014, Schulze_Wisotzki_2014,Sun_etal_2014}. Indeed, such a strong evolution in the slope (or the normalization) of the $M_{\bullet}-\sigma_*$ relation to $z\sim 1$ is inconsistent with the latest cosmological hydrodynamical simulations \cite[][]{Khandai_etal_2014,Sijacki_etal_2014,DeGraf_etal_2014}, or independent constraints from evolutionary models of the velocity dispersion function and accreted BH mass density \citep{Shankar_etal_2009}, which found no significant evolution in the $M_{\bullet}-\sigma_*$ relation.


\section{Conclusions}\label{sec:con}

We have presented a large, homogenous set of new host stellar velocity dispersion measurements in a uniformly-selected sample of quasars at $0.1<z<1$, with 46 quasars at $z>0.6$, a regime rarely explored in previous studies. The sample size and dynamic range in luminosity show significant improvement over the best samples studied previously \citep[e.g.,][]{Woo_etal_2006,Woo_etal_2008}. This work was made possible by the high S/N coadded spectra of a homogeneous quasar sample covering a wide luminosity and redshift range from the SDSS-RM project, the first major multi-object reverberation mapping project to date \citep{Shen_etal_2014}.

We decomposed the spectra into host and quasar components using the PCA method, and measured $\sigma_*$ and SE virial mass estimates based on the decomposed galaxy and quasar spectra, respectively. 

We found the first direct, conclusive evidence that BH mass correlates with $\sigma_*$ in quasars at $z>0.3$, such that quasar hosts with higher $\sigma_*$ have more massive BHs. Similar claims have previously been made, but were based on proxies for $\sigma_*$ that are much easier to measure in quasar spectra but with significant caveats \citep[such as the width of the \OII\ or \OIII\ line, e.g.,][]{Salviander_etal_2007,Salviander_Shields_2013}. This correlation holds to $z\sim 0.8$. We do not find significant evolution in this relation with our large sample and uniform analysis, up to $z\sim 1$. 

However, the $M_{\rm BH,vir}-\sigma_*$ relation is significantly shallower than the local $M_\bullet-\sigma_*$ relation (Figs.\ \ref{fig:msigma} and \ref{fig:msigma_evo}). We argued (e.g., Figs.\ \ref{fig:msigma_comp} and \ref{fig:sim_bias}) that this is most likely due to the statistical biases induced by the combination of the luminosity-threshold quasar selection and the uncertainties in the SE virial BH masses. The observed trend of the $M_{\rm BH,vir}-\sigma_*$ relation with luminosity (both within our sample and combined with other samples) appears to confirm this suggestion, but clearly more data and tests are needed. It is possible that there is still some real evolution to some extent. 

Our work represents a step forward in studies of the evolution of the $M_\bullet-\sigma_*$ relation, and demonstrates the potential and caveats of using quasars and virial BH mass estimates to probe the co-evolution of BHs and galaxies up to high redshift. Our SDSS-RM program has reached a depth and sample size that are difficult to surpass with traditional programs on current 6-10 m telescopes, but future installment of multi-object RM programs on 6-10 m telescopes or spectroscopy with 30~m-class telescopes will be able to provide even fainter, and hence less biased, quasar samples to measure host stellar velocity dispersion. With future direct RM-based BH masses for these quasars (the main goal of the SDSS-RM project), we can largely eliminate the luminosity-dependent bias in SE virial masses due to the explicit luminosity dependence of these mass estimators. It is unlikely that we will obtain a sample with RM masses and $\sigma_*$ measurements nearly as large as the one studied here. Nevertheless, these RM masses will help us understand better the systematics in these SE virial masses. With fainter quasar samples at high $z$, we can potentially recover the population of BHs that are undermassive relative to their hosts. With these two future advances, we will be able to better characterize the relation between BHs and host galaxies at high $z$, which in turn will improve our understanding of galaxy formation and BH feedback. 

\acknowledgements  We thank the anonymous referee for useful comments and suggestions, and Scott Tremaine and Andrew Benson for their insights. Support for the work of YS was provided by NASA through Hubble Fellowship grant number HST-HF-51314, awarded by the Space Telescope Science Institute, which is operated by the Association of Universities for Research in Astronomy, Inc., for NASA, under contract NAS 5-26555. LCH acknowledges support by the Chinese Academy of Sciences through grant No. XDB09030102 (Emergence of Cosmological Structures) from the Strategic Priority Research Program and by the National Natural Science Foundation of China through grant No. 11473002. WNB acknowledges support from NSF grant AST-1108604. BMP acknowledges support from the NSF through grant AST-1008882.


Funding for SDSS-III has been provided by the Alfred P. Sloan Foundation, the
Participating Institutions, the National Science Foundation, and the U.S.
Department of Energy Office of Science. The SDSS-III web site is
http://www.sdss3.org/.

SDSS-III is managed by the Astrophysical Research Consortium for the
Participating Institutions of the SDSS-III Collaboration including the
University of Arizona, the Brazilian Participation Group, Brookhaven National
Laboratory, University of Cambridge, Carnegie Mellon University, University
of Florida, the French Participation Group, the German Participation Group,
Harvard University, the Instituto de Astrofisica de Canarias, the Michigan
State/Notre Dame/JINA Participation Group, Johns Hopkins University, Lawrence
Berkeley National Laboratory, Max Planck Institute for Astrophysics, Max
Planck Institute for Extraterrestrial Physics, New Mexico State University,
New York University, Ohio State University, Pennsylvania State University,
University of Portsmouth, Princeton University, the Spanish Participation
Group, University of Tokyo, University of Utah, Vanderbilt University,
University of Virginia, University of Washington, and Yale University.

\end{document}